%% file: 0_main.tex
\documentclass{Interspeech2024}
\usepackage{boldline,multirow,multicol}
\usepackage{graphicx}
\usepackage{subfig}
\usepackage{pgfplots, mathtools}
\usepackage{xcolor}
\usepackage{cite}
\usepackage{footmisc}
\usepackage{authblk}
\usepackage{etoolbox}

\input{preamble}

\interspeechcameraready

\title{Speaker-Independent Acoustic-to-Articulatory Inversion through Multi-Channel Attention Discriminator}

\name[affiliation={1}]{Woo-Jin}{Chung}
\name[affiliation={1}]{Hong-Goo}{Kang}

\address{$^1$Dept. of Electrical \& Electronic Engineering, Yonsei University, South Korea}

\email{woojinchung@dsp.yonsei.ac.kr, hgkang@yonsei.ac.kr}

\keywords{Multi-phoneme duration analysis, Channel-wise attention, Electromagnetic articulography (EMA), Acoustic-to-articulatory inversion, Self-supervised learning}

\newcommand\blfootnote[1]{%
  \begingroup
  \renewcommand\thefootnote{}\footnote{#1}%
  \addtocounter{footnote}{-1}%
  \endgroup
}

\begin{document}
\maketitle

\input{1_abstract.tex}
\input{2_1_Introduction.tex}

\input{3_Architecture.tex}

\input{4_Experiments.tex}
\input{5_Conclusion.tex}
\

\clearpage
\bibliographystyle{IEEEtran}
\bibliography{longstrings,mybib}
\end{document}

%% file: preamble.tex
\definecolor{turquoise}{cmyk}{0.65,0,0.1,0.3}
\definecolor{purple}{rgb}{0.65,0,0.65}
\definecolor{dark_green}{rgb}{0, 0.5, 0}
\definecolor{orange}{rgb}{0.8, 0.6, 0.2}
\definecolor{red}{rgb}{0.8, 0.2, 0.2}
\definecolor{darkred}{rgb}{0.6, 0.1, 0.05}
\definecolor{blueish}{rgb}{0.0, 0.3, .6}
\definecolor{light_gray}{rgb}{0.7, 0.7, .7}
\definecolor{pink}{rgb}{1, 0, 1}
\definecolor{greyblue}{rgb}{0.25, 0.25, 1}
\definecolor{dark_gray}{rgb}{0.3, 0.3, 0.3}


%% file: 1_abstract.tex
\begin{abstract}

We present a novel speaker-independent acoustic-to-articulatory inversion (AAI) model, overcoming the limitations observed in conventional AAI models that rely on acoustic features derived from restricted datasets. 
To address these challenges, we leverage representations from a pre-trained self-supervised learning (SSL) model to more effectively estimate the global, local, and kinematic pattern information in Electromagnetic Articulography (EMA) signals during the AAI process.
We train our model using an adversarial approach and introduce an attention-based Multi-duration phoneme discriminator (MDPD) designed to fully capture the intricate relationship among multi-channel articulatory signals. 
Our method achieves a Pearson correlation coefficient of 0.847, marking state-of-the-art performance in speaker-independent AAI models.
The implementation details and code can be found online\blfootnote{This work was supported by the 'Alchemist Project' (Fully implantable closed-loop Brain to X for voice communication) funded By the Ministry of Trade, industry \& Energy (MOTIE, Korea), under Grant 20012355 and NTIS 1415181023.}\footnote{https://github.com/Woo-jin-Chung/Multi-Duration-Phoneme-Discriminator-Acoustic-to-Articulatory-Inversion}. 
\end{abstract}




%% file: 2_1_Introduction.tex
\begin{figure*}[!ht]
    \centering
    \includegraphics[width=\linewidth]{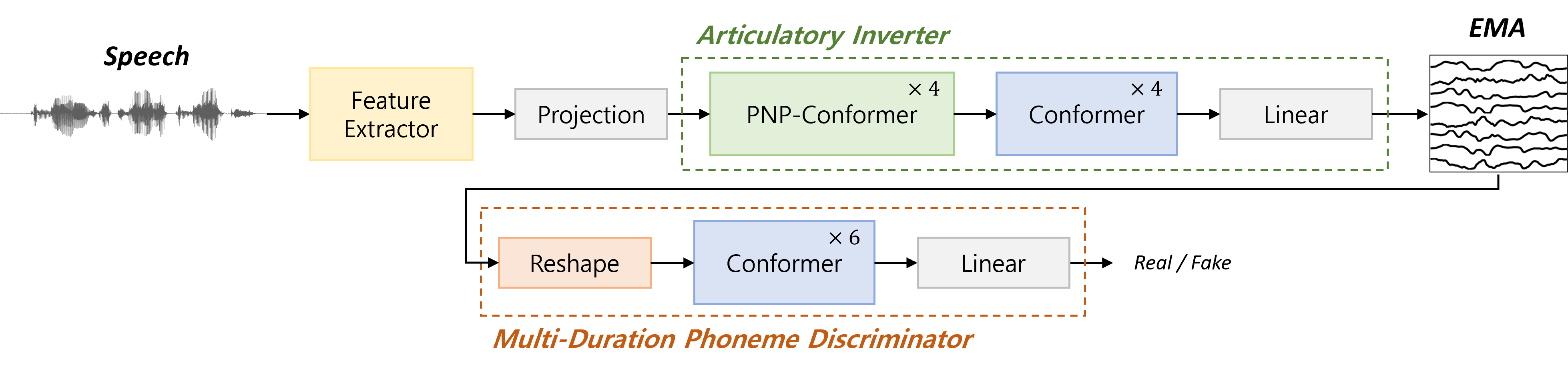}
    \vspace{-20pt}
    \caption{Illustration of the proposed model.}
    \label{fig_overall_architecture}
\end{figure*}

\vspace{-5pt}
\section{Introduction}

Articulatory movements, indicated by the positions of mouth components and measured through Electromagnetic Articulography (EMA), can be utilized as features in alternative communication methods for individuals with voice disorders.
There have been several approaches that utilize them for this purpose, such as articulation-to-speech (ATS)~\cite{chen2021ema2s}, silent speech interfaces (SSI)~\cite{denby2010silent} and brain-to-speech interfaces~\cite{anumanchipalli2019speech}.
However, obtaining articulatory data is often expensive, resulting in scarce data that poses significant challenges for models relying on it~\cite{illa2018low, maharana2021acoustic, wang2023two}. 
Consequently, acoustic-to-articulatory (AAI) modeling has emerged as a promising and cost-effective approach for extracting kinematic parameters from speech signals.
AAI aims to estimate articulatory signals directly from acoustic features~\cite{atal1978inversion}, offering a pragmatic solution to the limitations posed by data scarcity in traditional articulatory signal-based models.

Given the scarcity of data, researchers have explored various acoustic features related to the vocal tract that could effectively facilitate speaker-dependent AAI.
The most commonly used acoustic features in AAI include mel-frequency cepstral coefficients (MFCCs)~\cite{wang2022acoustic}, line spectral frequencies (LSF)~\cite{sivaraman2016vocal}, reflection coefficients (RC)~\cite{ghosh2010generalized}, and Log area ratio (LAR)~\cite{ghosh2010generalized}.
Building upon these acoustic features, data-driven AAI models have evolved with the advancement of deep learning techniques~\cite{sivaraman2016vocal, illa2019representation, wang2022acoustic, parrot2020independent, illa2019investigation}.
Recently, Transformer-based AAI models have emerged as state-of-the-art performers, particularly in speaker-dependent scenarios~\cite{udupa2021estimating, udupa2022streaming}.

Although much prior research focuses on speaker-dependent scenarios, there is a need for AAI for unseen speakers, from whom no articulatory data is available~\cite{ji2014speaker}.
Consequently, several works have introduced speaker-independent AAI models~\cite{wu2023speaker, bozorg2021autoregressive}, with \cite{wang2023two} recently achieving notable performance.
However, their method still obtains low inversion performance compared to speaker-dependent scenarios~\cite{shahrebabaki20b_interspeech}.
The main challenge lies in achieving sufficient generalization over unseen speakers with a simple data-driven approach, given the low feasibility of collecting large amounts of articulatory data for a significant number of speakers.

Regardless of speaker dependency scenarios, several prior works have attempted to transform input acoustic features into representations obtained from pre-trained self-supervised learning (SSL) models to address the data scarcity.
Since pre-trained SSL models are trained on large unlabeled speech corpora, they offer rich and robust acoustic representations that can be used in various downstream tasks~\cite{chen2022wavlm}. 
Representations extracted from speech SSL models have been shown to align with principles of speech production and articulatory gestures~\cite{cho2023evidence}.
Moreover, representations extracted from deeper layers of SSL models are known to be marginalized over speaker identity~\cite{chen22g_interspeech, pasad2021layer}, which suggests their potential for achieving good performance in tasks involving speaker-independent scenarios.

In this paper, we propose a novel acoustic-to-articulatory inversion (AAI) model over speaker-independent scenarios.
Our model is comprised of a Feature extractor, an Articulatory inverter, and a Multi-duration phoneme discriminator (MDPD).
We employ a pre-trained SSL model for the feature extractor and leverage the benefits of its representations to mitigate performance degradation caused by data limitation and speaker dependency.
We propose an Articulatory inverter based on the Conformer~\cite{gulati20_interspeech} architecture, aiming to capture both global and local features, incorporating it with periodic sensitive (PNP~\cite{chung23_interspeech}) Conformers to model detailed patterns of kinematic gestures.
Additionally, we use an adversarial training approach that utilizes a novel discriminator tailored to the multi-channel nature of articulatory features, the MDPD.
The discriminator is designed to learn the attentive relationship between the Electromagnetic articulography (EMA) channels for various phoneme durations as well as the relationships between the phoneme durations themselves.
To the best of our knowledge, this work represents the first attempt at improving speaker-independent AAI by explicitly utilizing characteristics of EMA signals for adversarial training.

Experimental results demonstrate that our proposed AAI model achieves state-of-the-art performance in speaker-independent scenarios, outperforming the prior best model in terms of Pearson correlation coefficient and root mean square error.

%% file: 3_Architecture.tex
\begin{figure}[!t]
    \centering
    \includegraphics[width=\linewidth]{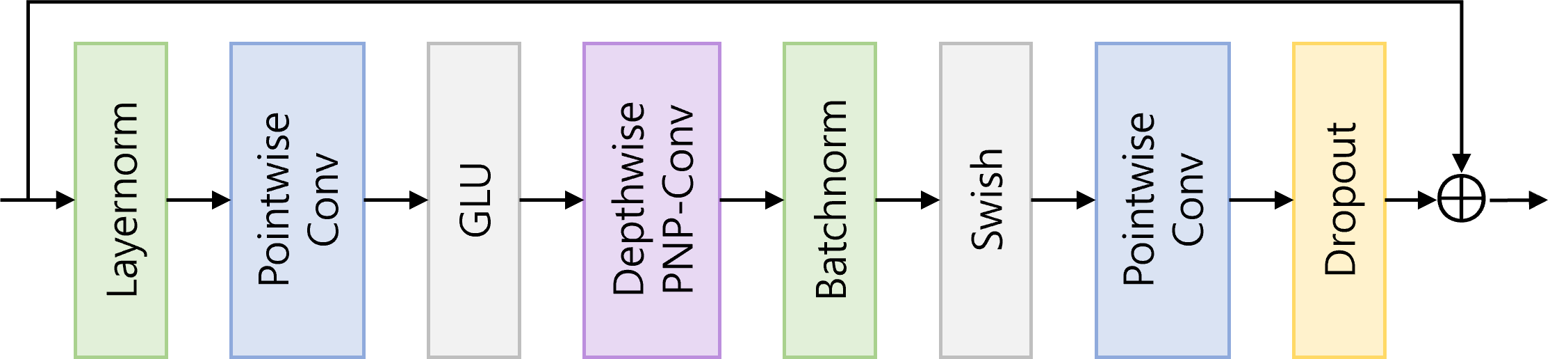}
    \caption{Illustration of PNP convolution module.}
    \label{fig_pnp_conv_module}
\end{figure}

\vspace{-4pt}
\section{Proposed model}
The overall structure of the proposed AAI model, as shown in Fig.~\ref{fig_overall_architecture}, comprises a Features extractor, an Articulatory inverter, and a Multi-duration phoneme discriminator (MDPD).
We employ a pre-trained SSL model as a Feature extractor due to the inadequacy of traditional acoustic features such as MFCCs in providing sufficient information for articulatory inversion.
The Articulatory inverter reverses the acoustic features extracted from the Feature extractor into articulatory signals. This process involves learning global, local, and kinematic patterns using Periodic sensitive (PNP) Conformers and Conformers.
MDPD captures the intricate relationship among multi-channel articulatory signals over phoneme periods by considering the characteristics of articulatory gestures, which serve as primitive phonological units.

\vspace{-2pt}
\subsection{Feature extractor}

Since a pre-trained SSL model is trained over a large corpus, it provides robust and comprehensive acoustic representations related to articulatory gestures.
In particular, representations derived from the deep layer of SSL models encapsulate acoustic properties that are invariant to speaker identity, thereby serving as viable alternatives to conventional acoustic features.
Given that WavLM~\cite{chen2022wavlm} has demonstrated superior performance compared to other speech SSL frameworks in various tasks, including the AAI task~\cite{cho2023evidence}, we utilize the pre-trained WavLM as the baseline of our Feature extractor.
Especially, we extract features from the 24-th layer of WavLM-Large\footnote{https://github.com/microsoft/unilm/tree/master/wavlm} which is the deepest layer, effectively mitigating the influence of speaker identity. 

\vspace{-2pt}
\subsection{Articulatory inverter}
\label{sec_articulatory_inverter}
\begin{figure}[!t]
    \centering
    \includegraphics[width=0.92\linewidth]{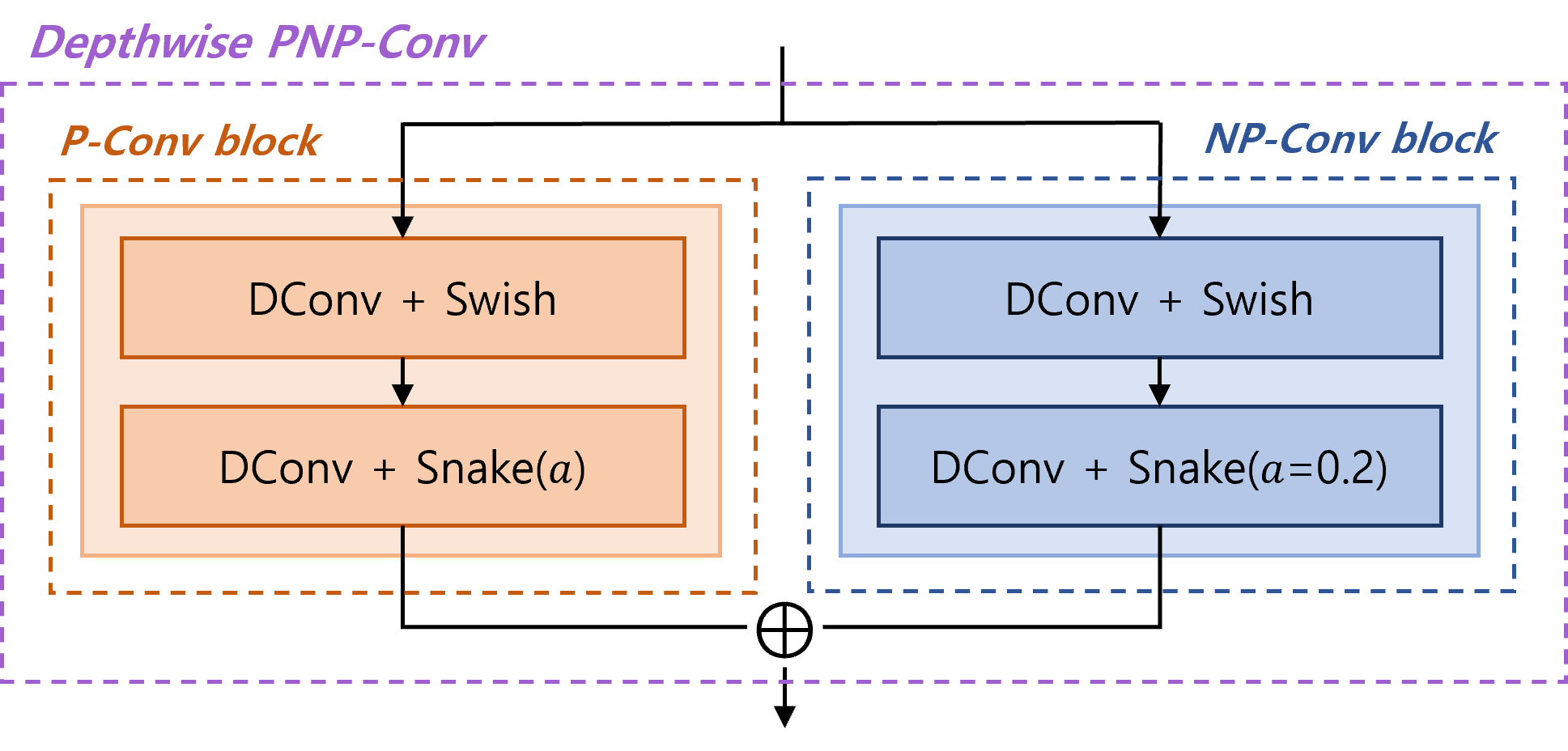}
    \caption{Illustration of depthwise PNP-convolution. Here, $a$ denotes the pre-fixed constant influencing the frequency range emphasized by the Snake activation function. DConv signifies the depthwise convolution operation.}
    \label{fig_depthwise_pnp_conv}
    \vspace{-10pt}
\end{figure}

The Articulatory inverter, comprising four PNP-Conformers~\cite{chung23_interspeech, gulati20_interspeech}, four Conformers~\cite{gulati20_interspeech}, and one fully connected layer, is structured to extract both global and local features, as well as the kinematic patterns conducive to the inversion process.
The Conformer architecture captures the global information of marginalized speaker identity and local details with Multi-head attention~\cite{vaswani2017attention} (MHA) and Convolutional neural network (CNN).
Moreover, the first four PNP-Conformer blocks enhance the capture of the kinematic patterns in features by substituting the conventional convolution module of the Conformer with the periodic-sensitive (PNP) convolution module~\cite{chung23_interspeech}, as illustrated in Fig.~\ref{fig_pnp_conv_module}.
Specifically, we replace the depthwise convolution with depthwise PNP-convolution, as illustrated in Fig.~\ref{fig_depthwise_pnp_conv}.
This modification incorporates the Snake activation function~\cite{ziyin2020neural}, which is designed to capture patterns across signals during model training.
In the P-Conv block of the depthwise PNP-Conv, we use the values (5, 7, 11, 13) for the $a$ factor of the Snake activation function in sequential order.
In all of our experiments, we apply a masking strategy where 15 $\%$ of the extracted embedding from the feature extractor is masked before inputting it into the PNP-Conformer blocks.
The masking strategy aligns with the approach used in wav2vec 2.0~\cite{baevski2020wav2vec}.

\begin{figure*}[!ht]
    \centering
    \includegraphics[width=0.97\linewidth]{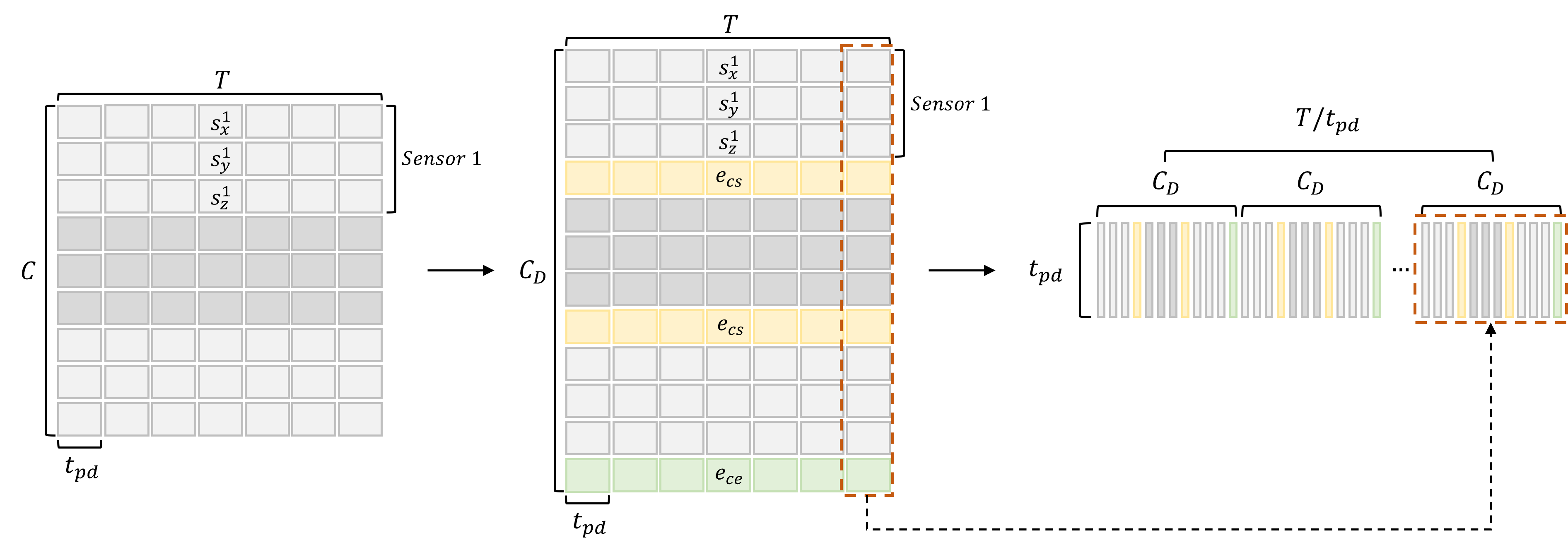}
    \vspace{-5pt}
    \caption{Illustration of the EMA reshaping process for sub-phoneme discriminators. 
            The left figure illustrates the EMA signals with $C$ EMA traces and a length of $T$.
            The middle figure shows the EMA signals after the addition of channel split embeddings ($e_{cs}$) and channel end embedding ($e_{ce}$).
            The Right figure displays the EMA signals for the MDPD input, reshaped to have a channel size corresponding to the phoneme duration ($t_{pd}$) and a length of $C_D\cdot(T/t_{pd})$.
            }
    \label{fig_ema_reshape}
\end{figure*}

\vspace{-2pt}
\subsection{Multi-duration phoneme discriminator}
\label{sec_mdpd}
To enhance the inversion performance through adversarial training, we develop a discriminator tailored to capture the unique characteristics of articulatory signals. 
Articulatory signals contain multi-channel information, recorded concurrently with speech production. 
These multi-channel articulatory gestures exhibit relationships among channels, as they are articulated simultaneously. Specifically, analysis based on phoneme periods is valuable, as they represent fundamental phonological units.

To capture the intricate relationship among articulatory multi-channels based on multi-phoneme durations, we propose the Multi-duration phoneme discriminator (MDPD).
MDPD comprises a mixture of sub-phoneme duration discriminators, each tailored to focus on attention across a singular phoneme duration $t_{pd}$.
Given that the average duration of a consonant phoneme falls within the range of 20 to 100 ms~\cite{crystal1988duration_c}, and vowels typically span from 70 to 180 ms~\cite{crystal1988duration_v}, we configure the phoneme durations to [60, 90, 100, 150, 180] ms to capture diverse implicit attention across each sub-phoneme duration discriminator.
During our experiments, we observed that establishing the phoneme duration as excessively short (e.g. 20 or 30 ms) for the sub-phoneme duration discriminator detrimentally impacts the model's inversion performance.
By utilizing scale dot-product self-attention with a slight reshaping of the EMA signal as input, MDPD could acquire the interconnected dynamics of the EMA gestures across phoneme durations, modeling to understand the relationship between EMA movements throughout phoneme periods.

The overall reshaping process is illustrated in Fig.~\ref{fig_ema_reshape}, where $s^n_x, s^n_y, s^n_z$ denote the $X$, $Y$, and $Z$ direction EMA signals of the $n^{th}$ sensor (TR, TB, TT, UL, LL, LI).
$e_{cs}$ denotes the learnable channel split embedding informing the model of the sensor changes and $e_{ce}$ denotes the learnable channel end embedding, indicating the end of the channel at a specific time.
$C$ denotes the total EMA traces which is 18 (6 sensors$\cdot$3 direction) and $C_D$ denotes the EMA channel after the addition of $e_{cs}$ and $e_{ce}$ which is 24 (18 traces + 5 $e_{cs}$ + 1 $e_{cs}$).
The reshaped EMA signal serves as input for 6 layers of Conformer blocks.
Due to the specially designed dimension size to match the phoneme duration $t_{pd}$, MDPD's Conformer blocks operate with single-head self-attentions to prevent unnecessary dimension splitting during the multi-head attention process.
We applied the same masking strategy to the reshaped EMA signals as explained in Section~\ref{sec_articulatory_inverter}.

\vspace{-2pt}
\subsection{Training criteria}
The entire model is trained in an end-to-end manner with a reconstruction loss ($\mathcal{L}_{recon}(I)$), adversarial losses ($\mathcal{L}_{adv}(D;I)$, $\mathcal{L}_{adv}(I;D)$), and a feature matching loss ($\mathcal{L}_{fm}(I;D)$).
The total loss is defined as,
\begin{equation}
\mathcal{L}_{I} = \mathcal{L}_{adv}(I;D) + \mathcal{L}_{recon}(I) + \mathcal{L}_{fm}(I;D),
\vspace{-2pt}
\label{eq_tot_gen_loss}
\end{equation}
\begin{equation}
\mathcal{L}_{D} = \mathcal{L}_{adv}(D;I),
\label{eq_tot_disc_loss}
\end{equation}
\begin{equation}
\vspace{+1pt}
\mathcal{L}_{total} = \mathcal{L}_{I} + \mathcal{L}_D,
\label{eq_tot_disc_loss}
\end{equation}
where $\mathcal{L}_{I}$ denotes the loss updated to the Articulatory inverter, $\mathcal{L}_{D}$ denotes the loss updated to the MDPD, and $\mathcal{L}_{total}$ denotes the total loss.

\vspace{2pt}
\noindent\textbf{Reconstruction loss.}
For the Articulatory inverter, we applied a reconstruction loss aimed at minimizing the discrepancy between the inverted EMA from the Articulatory inverter and the ground truth EMA.
We used the squared L2 norm (MSE) as the reconstruction loss.
It is defined as
\begin{equation}
\mathcal{L}_{recon}(I) = \mathbb{E}_{e,s}[||e-I(s)||^2_2].
\label{eq_recon_loss}
\end{equation}
where $I$ denotes the Articulatory inverter, $e$ denotes the ground truth EMA, and $s$ denotes the corresponding speech.

\vspace{2pt}
\noindent\textbf{Adversarial loss.}
We trained the model using adversarial training objectives, such as LS-GAN~\cite{mao2017least}, through our MDPD.
As explained in Section~\ref{sec_mdpd}, MDPD is designed to capture the attention of the multi-channel EMA over multi-phoneme durations by discriminating the output of the Articulatory inverter.
The MDPD is trained to classify the ground truth EMA as 1 and the inverted EMA from the Articulatory inverter as 0, leveraging EMA channel attention for discrimination.
The Articulatory inverter is trained to deceive the MDPD by estimating outputs that could be classified as a value of 1.
The adversarial loss is defined as
\begin{equation}
\mathcal{L}_{adv}(D;I) = \mathbb{E}_{(e,s)}[(D(e)-1)^2 + {D(I(s))}^2],
\label{eq_disc_loss}
\end{equation}
\begin{equation}
\mathcal{L}_{adv}(I;D) = \mathbb{E}_s[{(D(I(s))-1)}^2],
\label{eq_gen_loss}
\end{equation}
where $D$ denotes the MDPD.

\vspace{2pt}
\noindent\textbf{Feature matching loss.}
In our training approach, we incorporate the feature matching loss in addition to the adversarial loss to train the Articulatory inverter.
The feature matching loss compares the L1 distance of the feature maps of each layer in each sub-discriminator between the ground truth EMA and the inverted EMA.
The feature matching loss is defined as follows:
\begin{equation}
\mathcal{L}_{fm}(I;D) = \mathbb{E}_{e,s}\left[\sum_{i=1}^{T}\frac{1}{N_i}||D^i(e)-D^i(I(s))||_1\right],
\label{eq_fm_loss}
\end{equation}
where $T$ denotes the total number of layers in the MDPD ($D$), $N_i$ denotes the number of features from the $i$-th layer in the MDPD, and $D^i$ denotes the $i$-th layer of the MDPD.

%% file: 4_Experiments.tex
\vspace{-3pt}
\section{Experiments}
\subsection{Implementation details}\label{experimental_details}
\noindent \textbf{Dataset.}
The training and evaluation of the acoustic-to-articulatory inversion models are performed using the publicly available EMA dataset HPRC~\cite{tiede2017quantifying}.
The HPRC dataset has eight native American English speakers, four females (F01-F04) and four males (M01-M04), and 720 utterances respectively.
The dataset comprises eight sensors, tongue rear (TR), tongue blade (TB), tongue tip (TT), upper lip (UL), lower lip (LL), lower incisor (LI), mouth left (ML), and jaw left (JAWL).
In this work, we only used six locations (TR, TB, TT, UL, LL, LI) with X, Y, and Z directions.
The kinematic data undergoes min-max normalization and lowess smoothing to ensure stable training.
The speech data undergoes down-sampling to 16 kHz.
Eight models were built, each with eight speakers set to be unseen during the training process, aiming to address speaker-independent scenarios.

\vspace{2pt}
\noindent \textbf{Network configurations.}
All dimensions of the PNP-Conformer and Conformer modules utilized in both the Articulatory inverter and the MDPD are set to 256.
Every convolution layer used in the experiments employs a kernel size of 5 and a stride of 1.
The projection layer between the Feature extractor and the Articulatory inverter modifies the dimension of the representation using a single fully connected layer.

\vspace{2pt}
\noindent \textbf{Training details.}
All experiments are conducted with a batch size of 58 on a single A5000 GPU.
We initiate training of the MDPD after 28 epochs to reach the Nash equilibrium point, in accordance with the training technique employed in GANs, as detailed in \cite{salimans2016improved}.
We used an AdamW~\cite{loshchilov2018decoupled} optimizer along with a custom CosineAnnealingWarmRestarts~\cite{loshchilov2016sgdr} scheduler, with an initial learning rate of 0.0002 for both the Articulatory inverter and the MDPD.

\vspace{2pt}
\noindent \textbf{Evaluation protocols.}
We evaluated the articulatory inversion performance using the Pearson correlation coefficient (PCC) and Root mean square error (RMSE).
PCC is utilized to assess the linear correlation between the inverted and ground truth kinematics, while RMSE quantifies the average difference between them. 

\vspace{-5pt}
\subsection{Results}
\input{tables/comparison}
We first evaluate the effectiveness of our proposed AAI model by benchmarking its articulatory inversion performance against the state-of-the-art model in speaker-independent scenarios~\cite{wang2023two} (SI-SOTA).
Since SI-SOTA only utilized sensor locations in the X and Z directions, representing kinematic movements from posterior to anterior and inferior to superior, we additionally trained our proposed model using the same experimental data setting. 
The newly trained proposed model, referred to as "Proposed$\_$xz" as shown in Table~\ref{tab_comparison}.
The "Proposed$\_$xz" model demonstrated superior articulatory inversion performance compared to SI-SOTA, achieving values of 0.848 for PCC and 2.489 for RMSE.
However, our proposed model still shows slightly lower performance than the state-of-the-art model in speaker-dependent scenarios~\cite{shahrebabaki20b_interspeech} (SD-SOTA).

\vspace{2pt}
\noindent \textbf{Ablation study.}
To further investigate the individual contributions of the MDPD and each module in the Articulatory inverter, we conducted an ablation study, as shown in Table~\ref{tab:ablation_study}.
We first assessed the effectiveness of using the SSL representation, referred to as 'MFCC input'. 
The result demonstrates a significant improvement in utilizing SSL representation for speaker-independent AAI compared to traditional acoustic features.

We then replaced the Articulatory inverter architecture with various architectures to evaluate the effectiveness of learning both global and local information using Conformers and modeling periodic patterns with PNP-Conformers.
We replaced the Articulatory inverter with standard Conformers (w/o. PNP), transformers (w/o. Local info.), convolution neural network (w/o. Global info.), and Multilayer perceptron (MLP).
As shown in Table~\ref{tab:ablation_study}, 'w/o. PNP' outperforms all other models ('w/o. Local info.', 'w/o. Global info.', and 'MLP'). This suggests that modeling the Articulatory inverter based on the Conformer architecture, to capture both global and local information, mainly contributes to the inversion performance in speaker-independent scenarios. The comparison between the proposed model and 'w/o. PNP' demonstrates the effectiveness of the PNP module in capturing kinematic patterns.

We also conducted an experiment removing the MDPD and adversarial training, relying solely on the reconstruction loss. 
This setup, referred to as 'no MDPD', allowed us to assess the effectiveness of our proposed MDPD.
In the 'w/o. MDPD' setup, the absence of MDPD results in the failure to capture EMA channel attention observed over and across phoneme durations. This led to a significant performance degradation compared to the proposed AAI model.
These results highlight the crucial role played by our designed Articulatory inverter and proposed MDPD in improving the articulatory inversion performance in our speaker-independent scenario.

\input{tables/ablation_study}

%% file: tables/comparison.tex
\begin{table}[]
\centering
\caption{Results of PCC and RMSE on the HPRC dataset.}
\label{tab_comparison}
\vspace{-5pt}
\resizebox{\columnwidth}{!}{
\setlength{\tabcolsep}{3pt}
\renewcommand{\arraystretch}{1.2}
\begin{tabular}{l||cccccccc|cc}
\toprule
\multirow{2}{*}{\textbf{Models}} & \multicolumn{8}{c|}{\qquad \textbf{Speakers} (PCC $\uparrow$)}                           & \multicolumn{2}{c}{\textbf{Total}} \\ 
                        & F01   & F02   & F03   & F04   & M01   & M02   & M03   & M04   & \;\,PCC $\uparrow$        & \;\,RMSE $\downarrow$       \\ \midrule
SD-SOTA~\cite{shahrebabaki20b_interspeech}  & 0.945 & 0.887 & 0.861 & 0.933 & 0.902 & 0.893 & 0.860 & 0.875 & \textbf{0.895}       & \textbf{1.216}       \\ \midrule
SI-SOTA~\cite{wang2023two}  & -     & -     & -     & -     & -     & -     & -     & -     & 0.810       & 2.537       \\ \midrule
Proposed$\_$xz            & 0.863 & 0.852 & 0.874 & 0.901 & 0.834 & 0.832 & 0.812 & 0.814 & \textbf{0.848}       & 2.489       \\ \midrule
\textbf{Proposed}                & 0.862 & 0.854 & 0.878 & 0.893 & 0.841 & 0.826 & 0.801 & 0.819 & 0.847       & \textbf{2.467}       \\ \bottomrule
\end{tabular}
}
\vspace{-10pt}
\end{table}

%% file: tables/ablation_study.tex
\begin{table}[]
\centering
\caption{Results of average PCC and RMSE for ablation studies conducted on MDPD and each module in the Articulatory Inverter. The parameter size of MDPD is not included.}
\label{tab:ablation_study}
\vspace{-5pt}
\resizebox{0.8\columnwidth}{!}{
\begin{tabular}{l||c|c|c}
\toprule
\multirow{2}{*}{\textbf{Models}} & \multirow{2}{*}{\begin{tabular}[c]{@{}c@{}}\textbf{Params.}\\ (\textbf{M})\end{tabular}} & \multirow{2}{*}{\;\, \textbf{PCC} $\uparrow$ } & \multirow{2}{*}{\;\, \textbf{RMSE} $\downarrow$} \\
                        &                                                                        &                      &                       \\ \midrule
Proposed                & 12.9                                                                   & 0.847                & 2.871                 \\ 
MFCC input              & 12.6                                                                   & 0.642                & 3.357                 \\ 
w/o. PNP                & 12.6                                                                   & 0.839                & 3.315                 \\ 
w/o. Local info.        & 12.6                                                                   & 0.829                & 2.909                 \\ 
w/o. Global info.       & 12.5                                                                   & 0.696                & 3.119                 \\ 
MLP                     & 12.7                                                                   & 0.713                & 3.057 \\ 
w/o. MDPD               & 12.9                                                                  & 0.811               & 4.390                 \\ \bottomrule
\end{tabular}
}
\vspace{-10pt}
\end{table}


%% file: 5_Conclusion.tex
\vspace{-5pt}
\section{Conclusions}
In this paper, we proposed a novel speaker-independent acoustic-to-articulatory inversion (AAI) model that utilizes representations from a speech self-supervised learning model and adversarial training to effectively learn the transformation between speech to Electromagnetic Articulography (EMA) signals.
Our proposed model captures global, local, and kinematic pattern information from an Articulatory inverter and learns the EMA channel attention over multiple phoneme durations from a novel Multi-duration phoneme discriminator (MDPD).
Experimental results demonstrated that our AAI model outperformed the previous state-of-the-art model in speaker-independent settings.
We also conducted ablation studies to demonstrate the contributions of each module to its overall performance.